\documentclass[conference]{IEEEtran}
\IEEEoverridecommandlockouts
\usepackage{cite}
\usepackage{amsmath,amssymb,amsfonts}
\usepackage{algorithmic}
\usepackage{graphicx}
\usepackage{textcomp}
\usepackage{xcolor}
\def\BibTeX{{\rm B\kern-.05em{\sc i\kern-.025em b}\kern-.08em
    T\kern-.1667em\lower.7ex\hbox{E}\kern-.125emX}}

\definecolor{color_yuqi2}{RGB}{234, 135, 47}

\usepackage{stfloats}
\usepackage{subfig}
\usepackage{graphicx}

\begin{document}

\title{Machine Learning-Assisted Distribution System Network Reconfiguration Problem
\thanks{This work was partly authored by the National Renewable Energy Laboratory, operated by Alliance for Sustainable Energy, LLC, for the U.S. Department of Energy (DOE) under Contract No. DE-AC36-08GO28308. Funding for this work was provided by the U.S. Department of Energy Office of Electricity under the Microgrid program. The views expressed in the article do not necessarily represent the views of the DOE or the U.S. Government. The U.S. Government retains, and the publisher, by accepting the article for publication, acknowledges that the U.S. Government retains a nonexclusive, paid-up, irrevocable, worldwide license to publish or reproduce the published form of this work or allow others to do so for U.S. Government purposes.
}
}

\author{
\IEEEauthorblockN{Richard Asiamah}
\IEEEauthorblockA{\textit{Georgia Institute of Technology} \\
Atlanta, GA, USA \\
asiamah@gatech.edu}
\and
\IEEEauthorblockN{Yuqi Zhou}
\IEEEauthorblockA{\textit{National Renewable Energy Laboratory} \\
Golden, CO, USA \\
yuqi.zhou@nrel.gov}
\and
\IEEEauthorblockN{Ahmed S. Zamzam}
\IEEEauthorblockA{\textit{Ascend Analytics} \\
Boulder, CO, USA \\
azamzam@ascendanalytics.com}
}

\maketitle

\begin{abstract}
High penetration from volatile renewable energy resources in the grid and the varying nature of loads raise the need for frequent line switching to ensure the efficient operation of electrical distribution networks. Operators must ensure maximum load delivery, reduced losses, and the operation between voltage limits. However, computations to decide the optimal feeder configuration are often computationally expensive and intractable, making it unfavorable for real-time operations. This is mainly due to the existence of binary variables in the network reconfiguration optimization problem. 
To tackle this issue, we have devised an approach that leverages machine learning techniques to reshape distribution networks featuring multiple substations. This involves predicting the substation responsible for serving each part of the network.
Hence, it leaves simple and more tractable Optimal Power Flow problems to be solved. This method can produce accurate results in a significantly faster time, as demonstrated using the IEEE 37-bus distribution feeder. Compared to the traditional optimization-based approaches, a feasible solution is achieved approximately ten times faster for all the tested scenarios. 
\end{abstract}

\begin{IEEEkeywords}
Microgrids, deep neural networks, optimal power flow
\end{IEEEkeywords}

\section{Introduction}
Grid operators solve the optimal power flow (OPF) problem to determine the most economical generation dispatch path to meet all the load demands while simultaneously ensuring all constraints are met.
In distribution networks with flexible reconfiguration capabilities, this is also accompanied by a problem in determining the optimal topology to reduce operation costs further and relieve congestion~\cite{distribution_reconfiguration}.
Integrating more renewable energy resources into the grid has raised the need for more frequent solutions to these problems to cater to uncertainty surrounding the intermittent nature of generation sources such as wind and solar. Remote-controlled switches are strategically placed along switchable lines to enable real-time line control, leading to greater flexibility in network configuration~\cite{taylor2023managing, zhou2024equitable}.
However, the biggest challenge is the complexity of performing these computations. The presence of binary variables representing the status of switchable lines adds significant complexity to the optimization problem, where the number of possible configurations increases exponentially with the number of switchable lines. Hence, solving such problems is often computationally intensive, making real-time grid operations difficult~\cite{robust}. 

Recent literature has started utilizing artificial intelligence techniques to enhance grid operations. Machine learning-based approaches have been proposed to solve OPF problems without relying on traditional solvers to compute solutions~\cite{ac_with_nn, joint_chance, dc_with_nn, dc_sets}. Many of these take different techniques, including learning a good starting point for AC OPF, learning the active set of constraints for the DC OPF problem~\cite{dc_sets}, or simply taking advantage of large quantities of measurement data that are being generated but are not yet being fully utilized to predict the solutions to the OPF problems~\cite{ac_with_nn}.
Besides optimal power flow solutions, neural network learning-based approaches have also been applied in many different applications in power systems optimization. For instance, in \cite{unit_commitment}, the authors show that compact neural networks can be exactly transformed into mixed integer linear programs and embedded inside computationally challenging optimization problems such as the AC Unit Commitment problem while \cite{SE_with_nn} tackles the Distribution System State Estimation problem using a neural network based learning model.
The work in \cite{yin2020data} proposed a real-time online data-driven distribution network reconfiguration (3DNR) method. They demonstrated that convolutional neural networks could solve the real-time distribution network reconfiguration (DNR) problem without power flow calculation. They tested the feasibility and effectiveness of the proposed method on different test cases. However, training neural networks to learn the reconfiguration problem solutions directly requires a voluminous amount of training data, and it is often a complex learning task for neural network models.

\begin{figure}[t!] 
  \centering
  \includegraphics[trim=0cm 0cm 0cm 0.3cm,clip=true, width=0.30\textwidth]{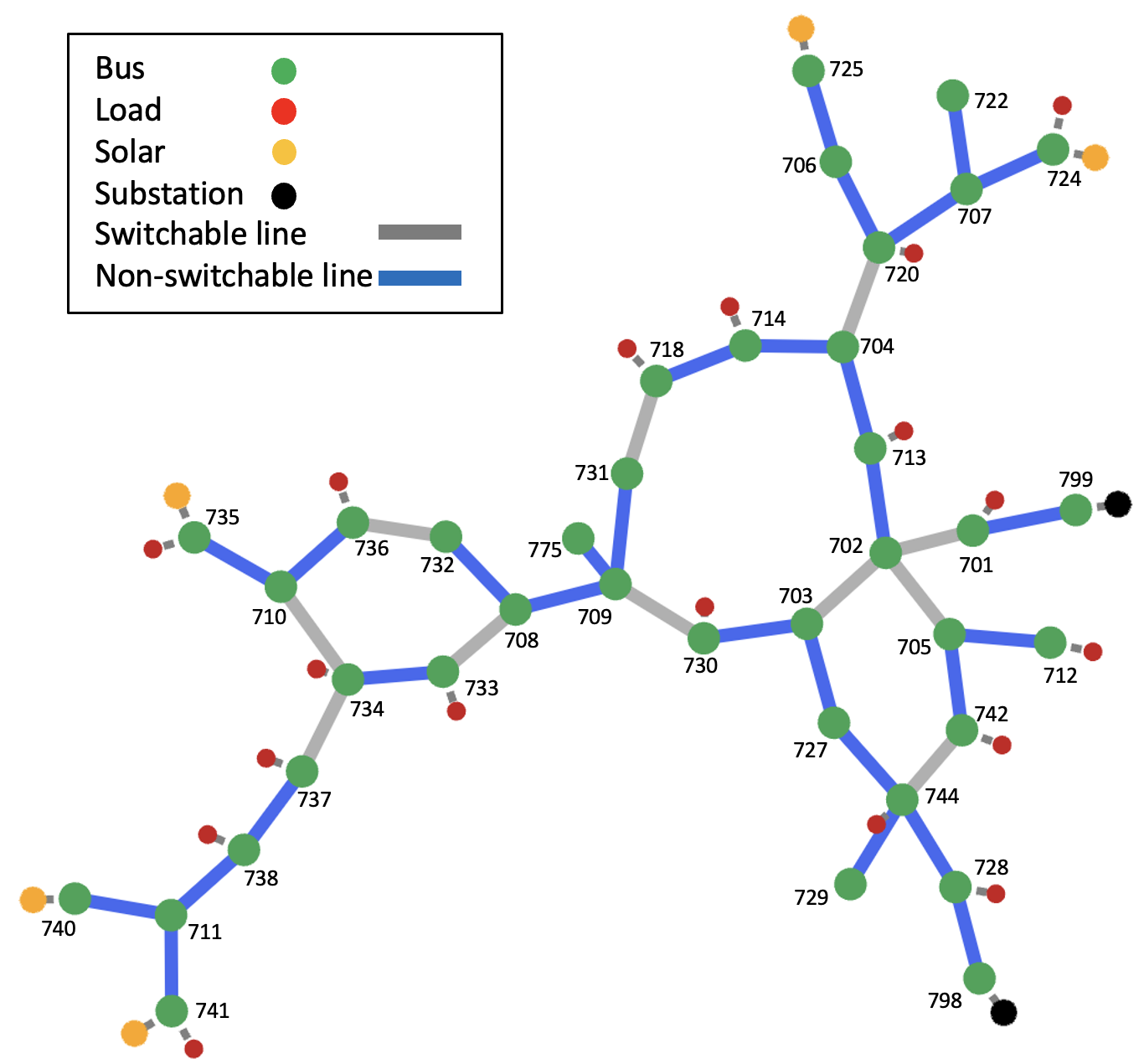}
  \caption{The modified IEEE 37-Bus system with flexible network topology configuration.}
  \label{fig:37_bus}
\vspace*{-2mm}
\end{figure}

\begin{figure}[t!] 
  \centering
  \includegraphics[scale = 0.2]{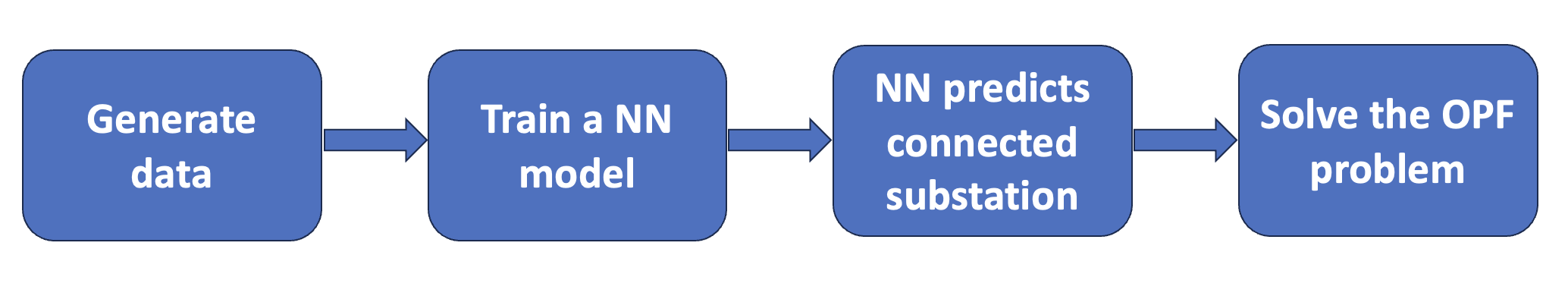}
  \caption{Overall architecture of the machine learning framework}
  \label{fig:workflow}
\vspace*{-3mm}
\end{figure}

In this paper, we seek to learn a substation assignment problem. That is, we use neural networks to estimate which substation is feeding energy to each component in the network, avoiding the long solving time associated with the microgrid reconfiguration problems. We run and solve our computationally demanding optimization problem several times offline to generate enough data. This data would be in the form of load values and PV generated at the individual solar plants as inputs and the corresponding optimal network configuration or path as the output. The route is determined by using the substation to which each load block in the system is connected and deducing the shape of the network from this information (see Fig.~\ref{fig:37_bus}). 
Depending on the load profile and the generation available at each PV station, the network topology is optimized to minimize total operating costs. The load and PV inputs are varied to generate different instances with corresponding optimal configurations. The connected substations can then be inferred from the optimal configuration of the network. The data collected are then used to train a neural network by learning the mapping from the value of the loads and the generation of PV generators to the substation to which each component is connected. This way, the optimal network topology can be uniquely determined. 
The overall learning architecture of the proposed approach is demonstrated in Fig.~\ref{fig:workflow}.
To prevent predictions with infeasible solutions, physically unrealistic configurations are adjusted to the closest viable alternatives, while voltage constraints can be relaxed to allow for occasional voltage violation if necessary.

The structure of the paper is as follows: Section II focuses on the optimal network reconfiguration problem formulation. Section III provides further insights into the steps and processes taken in learning the data to enhance the accuracy of the predictions. Section IV validates the proposed method using the modified IEEE 37-bus distribution feeder, and Section V concludes the paper.

\section{Problem Formulation}
This section mainly introduces the optimization formulation of the optimal network reconfiguration problem.  
We approach the problem from the perspective of a power system operator whose primary goal is to ensure all load demands are met at minimal operation costs.
We assume that operators would use this method for day-to-day operations to decide on the best configuration of the distribution network and the optimal dispatch of flexible resources within the grid. 
Let $\mathcal{G}$, $\mathcal{B}$, and $\mathcal{L}$ denote the sets of generators, buses, and branches, respectively, in the network. The number of loads and solar PV units are represented by $\mathcal{D}$ and  $\mathcal{S}$, respectively. The apparent power generation of each generator is represented as $S_{g}^G := P_{g}^G +i Q_g^G$, for $g \in \mathcal{G}$. Similarly, the load demand at bus $i\in \mathcal{B}$ is denoted by $S_i^L := P_i^L +iQ_i^L$, and the apparent power flow in each branch $l \in \mathcal{L}$ from bus $i$ to bus $j$ is given as $S_{ij}:=P_{ij} + i Q_{ij}$, respectively. The voltage at every bus is $V_i \in\mathbb{C}^{|\Phi_i|}$, $i \in \mathcal{B}$, where $\Phi_i$ is the set of phases at bus $i$. Then, we define $v_i := V_i V_i^H$ to denote the matrix of squared voltages. Below, we present the constraints to this problem and the complete optimization problem.

\subsubsection{Generation Limits}
The generator limits are given by 
\begin{align}
    \underline{P}_{g} \leq P_{g}^G \leq \overline{P}_{g} \qquad  \forall g \in \mathcal{G} \label{gen_constraints}\\
    \underline{Q}_{g} \leq Q_{g}^G \leq \overline{Q}_{g} \qquad  \forall g \in \mathcal{G} \label{gen_constraints2}
\end{align}
which enforces that the value of the active and reactive power produced by the generators is within the maximum and minimum limits. For solar units, the lower bounds are zero, while the upper bounds are the available power based on solar irradiance.

\subsubsection{Line Flow and Bus Voltage Limits}
The power flow along the line $l \in \mathcal{L}$ must not exceed the upper and lower line flow limits. The same applies to the voltage at each of the buses, $i \in \mathcal{B}$. They are constrained to stay between predefined limits for optimal operation of the network.
\begin{align}
    \underline{S}_{ij} \leq \text{diag}(S_{ij}) \leq \overline{S}_{ij} \qquad  \forall l \in \mathcal{L} \label{line_power_limits}\\
    \underline{V}_{i}^2 \leq \text{diag}(v_{i}) \leq \overline{V}_{i}^2\qquad  \forall i \in \mathcal{B} \label{voltage_limits}
\end{align}

\subsubsection{Linearized Power Flow Model}
Our problem is solved using the Linearized Power Flow approximation (LPF) formulation, a linear approximation of the AC power flow model obtained by assuming negligible line losses and balanced voltages across all nodes. In~\cite{lpf}, LPF is shown to be a good estimate of apparent power and voltages in a network. It is shown to provide much better results than the DC power flow model since it does not assume constant voltage or ignore reactive power.

Let $s_i := S_{i}^G - S_i^L$ denote the power injection at the bus, $i \in \mathcal{B}$ for all phases in $\Phi_i$. The impedance across a branch from bus $i$ to $j$ is denoted by $z_{ij}$, and we introduce a new variable, $\Lambda_{ij}~:=~diag(S_{ij})$. To this end, the power flow equations can be expressed as:
\begin{align}
    \sum_{i:i\rightarrow j}\Lambda_{ij}+s_{j}=\sum_{k:j\rightarrow k}\Lambda_{jk}^{\Phi_{j}},\quad \qquad j\in {\cal B}, \\ \label{lpf_1}
    S_{ij}=\gamma^{\Phi_{ij}} {\rm diag} (\Lambda_{ij}),\qquad \qquad \quad i\rightarrow j, \\ \label{lpf_2}
    v_{j}=v_{i}^{\Phi_{ij}}-S_{ij}z_{ij}^{H}-z_{ij}S_{ij}^{H},\qquad i\rightarrow j, 
\end{align}
for which $\alpha$ and $\gamma$ are defined as:
\begin{align}
\alpha := e^{-j2\pi/3} \quad \text{and} \quad
\gamma := \begin{bmatrix} 1 & \alpha^2 & \alpha \\ \alpha & 1 & \alpha^2 \\ \alpha^2 & \alpha & 1 \end{bmatrix}.
\end{align}

\subsubsection{Topology Constraints}
In solving this problem, the network should be operating in a radial structure, i.e., the network should be arranged in a tree structure or a collection of trees where each substation is the root of a tree. Additionally, no islands should be formed in the isolation of substations. These constraints are modeled using PowerModelsONM.jl software package \cite{ONM}. For brevity, we denote these constraints as 
\begin{align}
    \label{eq:radiality}
    {\bf A} {\bf x} = {\bf b} 
\end{align}
where ${\bf x}$ is a vector collecting the binary variables denoting the status of switchable lines, and ${\bf A}$ and ${\bf b}$ are the coefficients of the constraints determined based on the multi-commodity flow formulation for radiality. For further details, we refer the reader to \cite{ONM}. 

Given all the constraints described above, the optimal reconfiguration problem can be formulated as follows:
\begin{subequations}
\begin{align}
\text{Minimize:} \quad & \sum_{g \in \mathcal{G}} C_g(P_g)  \label{mld_problem}  \\ 
\text{Subject to:} \quad & \eqref{gen_constraints}-\eqref{eq:radiality}
\end{align}
\end{subequations}
where $C_g$ is a convex (typically quadratic or linear) cost function for capturing active power generation. For simplicity, we use a linear cost function to minimize the amount of power taken from the generators in the network. The resulting problem constitutes a mixed-integer linear program (MILP), which can be efficiently solved by optimization solvers such as HiGHS, GLPK, and CPLEX.

\section{Learning Optimal Reconfiguration Solutions}
This section discusses the methods used to generate the data sets, train our neural network to ensure high accuracy and obtain a feasible solution for each test case.

\subsection{Data Collection} 
The optimization problem was solved multiple times to generate feasible power flow solutions in which all thermal, voltage, and generation limits were satisfied. Real datasets of loads and PV generation were used to estimate the parameters of a normal distribution that best fits the variation in the data. 
During each iteration, load, and PV data were randomly sampled from a normal distribution to solve the problem. After that, the values of active loads, PV generation at each bus, and the substation to which the individual load blocks were connected were recorded. The loads varied between $80\%$ and $120\%$ of the nominal values, and the solar generation varied from $0$ to $100\%$ of the rated values. This range of distribution was chosen to maintain feasibility while increasing the richness of the data.
A total of 5000 test cases were run offline, and these values were stored as training data for our neural network.

\subsection{Training the Neural Network Model}
The neural networks being trained are required to learn the substation to which each load block is connected. A load block in an electrical network is any collection of buses, with or without loads, that cannot be physically separated from each other. The topology constraint in our problem ensures that each load block will be connected to \emph{only one} substation, and hence, our neural network has an output for each load block denoting which substation it is connected to. When there are only two substations, the output of the neural network can be just binary or can be categorized with a one-hot embedding representation if there are more than two substations.

\begin{figure}[t!] 
  \centering
  \includegraphics[scale = 0.18]{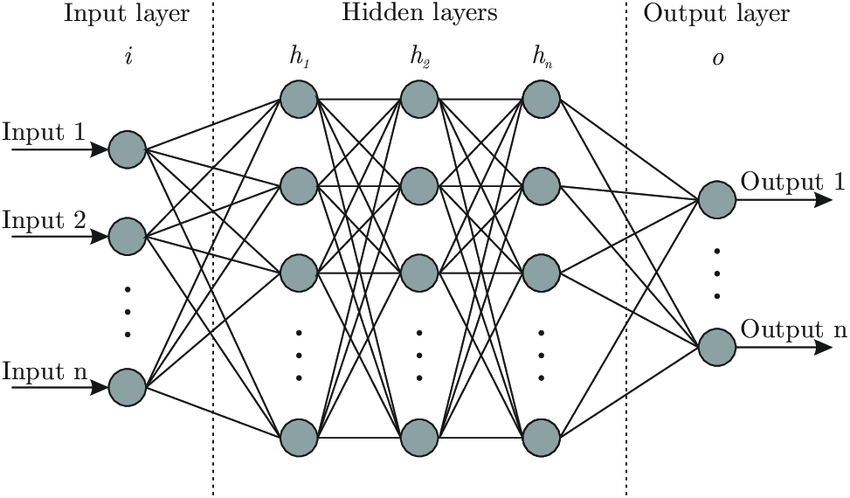}
  \caption{The deep neural network framework for learning the optimal network reconfiguration}
  \label{fig:nn}
\vspace*{-3mm}
\end{figure}

To accurately approximate the mapping, we utilize a deep neural network model with four layers. The size of the input of the neural network is $\lvert \mathcal{D+S} \rvert$, representing the number of active load demand at the load buses plus the number of solar panels. The number of outputs is the number of load blocks in the network. Fig. \ref{fig:nn} depicts the architecture of our approach's deep neural network learning model. The first three layers utilized a ReLU activation, and the last used a sigmoid activation function.  

Finally, the model is trained using the classical empirical loss minimization formulation, which is specifically tailored for binary classification problems. The binary cross-entropy loss is represented by the formulation below: 
\begin{align}
    L(y, \hat{y}) = - (y \log(\hat{y}) + (1 - y) \log(1 - \hat{y}))
\end{align}
where $y$ and $\hat{y}$ represent the ground truth array and the predicted array, respectively.
Initial efforts to create neural networks were inefficient due to high levels of overfitting. 
To avoid this, the following accuracy-enhancing techniques were employed to prevent the model from learning repetitive behaviors and help improve learning accuracy:  

\subsubsection*{Normalization}
The concept of normalization was first introduced in \cite{norm}. This technique is implemented by transforming the input to have a mean of zero and a standard deviation of one. Normalizing our data was essential, especially considering the difference in the range of values for our active load demands and PV generations. Normalizing the training data makes it more consistent, resulting in higher training rates and shorter solving time.

\subsubsection*{Regularization}
The simplest and perhaps most common regularization method is to add a penalty to the loss function proportional to the size of the weights in the model. Given the nature of the data, we used the $L_2$ regularization or the Ridge regularization on all layers, with the penalty factor set at 0.001.


\subsubsection*{Dropout}
The term dropout refers to dropping out units in some layers of the neural network to prevent overfitting or these units being learned multiple times~\cite{srivastava2014dropout}. Dropping out particular neurons temporarily takes out all its incoming and outgoing connections in the neural network to not affect the state of the learning process. In training our model, dropout was applied to the first three layers with a probability of any element to be dropped out set to 0.25.

\subsubsection*{Data Augmentation}
Generally, models' performance improves with increasing amounts of training data. The more the training data, the higher the probability of learning accuracies and predictive capabilities. A solution explored to address this is data augmentation. In this work, we increased the variation in the training data by including ``random noise'' in the input parameters of loads and PV. These newly generated points were added by randomly generating noise values from a normal distribution with a mean of 0 and a standard deviation of 1. This noise is added to the input, but the output value is still assumed to be the same.

\subsection{Validation and Testing}
The original data is divided into training, testing, and validation sets to achieve the best possible performance of the models. In our work, the total dataset was divided into three portions: 90\% for training, 5\% for validation, and the remaining 5\% as testing datasets. During the training process, validation and testing datasets are held back and reserved from training. Validation datasets are used to estimate the model's performance during training and aid in preventing overfitting by using early stopping. The final accuracy of the model is determined by comparing the prediction of the model over the testing dataset with the original values.

The neural network model predicts which substation provides power to each load block. This prediction information can be used to determine the configuration of the network and, hence, the status of each switch in the network. The optimal power flow problem is then solved with the status of all the switchable lines fixed.

\section{Numerical Results}
In this section, we describe the networks used in our experiments and highlight the observations made from simulations. We employ a standard distribution network to generate training data and use that to train a neural network. Once successfully trained, we use the neural network's predictions to set the network topology of the same network and proceed to solve the resulting OPF problem. We present the results of the methods used and assess the accuracy of the model.

\subsection{Implementation}
The optimal reconfiguration and OPF formulation of the problems were formulated and implemented in the Julia Language~\cite{julia} using JuMP~\cite{dunning2017jump}. 
The PowerModelsONM package~\cite{ONM} was utilized to model the reconfiguration problem. The optimization problems were solved using the HiGHS optimizer on a machine with a 12-thread Mac-M2 chip @.32GHz and 16 GB memory.
The deep neural networks' training, validation, and testing were implemented in Python with the Scikit-Learn package.

\subsection{Test Networks}
This method was implemented on the IEEE 37-bus test network. Slight modifications were made to the system, including a new substation at node 798, declaring 9 of the lines in the standard test case as switchable, and including 4 new switchable lines connecting different buses to eliminate the network's inherent radial configuration. Additionally, 5 solar PV units were added and assigned to random buses in the network. 
These changes also demonstrate how the unpredictable and random nature of solar energy generation adds uncertainty to the power grid, highlighting the necessity of network reconfiguration.
Details of the network can be found in TABLE~\ref{tab:my_label}.

\begin{table}[t!]
    \centering
    \caption{Breakdown of Components in the Modified 37-bus Test Case}
    \begin{tabular}{|c|c|c|} \hline 
           &\textbf{Original Network}&  \textbf{Modified Network}\\ \hline 
         Buses & 39 &  40\\ \hline 
         Lines & 36 &  41\\ \hline
         Loads & 30 &  30\\ \hline 
         PV stations & 0 &  5\\ \hline 
         Substations & 1 &  2\\ \hline 
         Switchable lines & 0 &  13\\\hline
    \end{tabular}
    \label{tab:my_label}
\vspace*{-2mm}
\end{table}

\begin{figure*}[t!]
\centering 
   \subfloat[Scenario 1]{
      \includegraphics[trim=0cm 0cm 0cm 0cm,clip=true, width=0.26\textwidth]{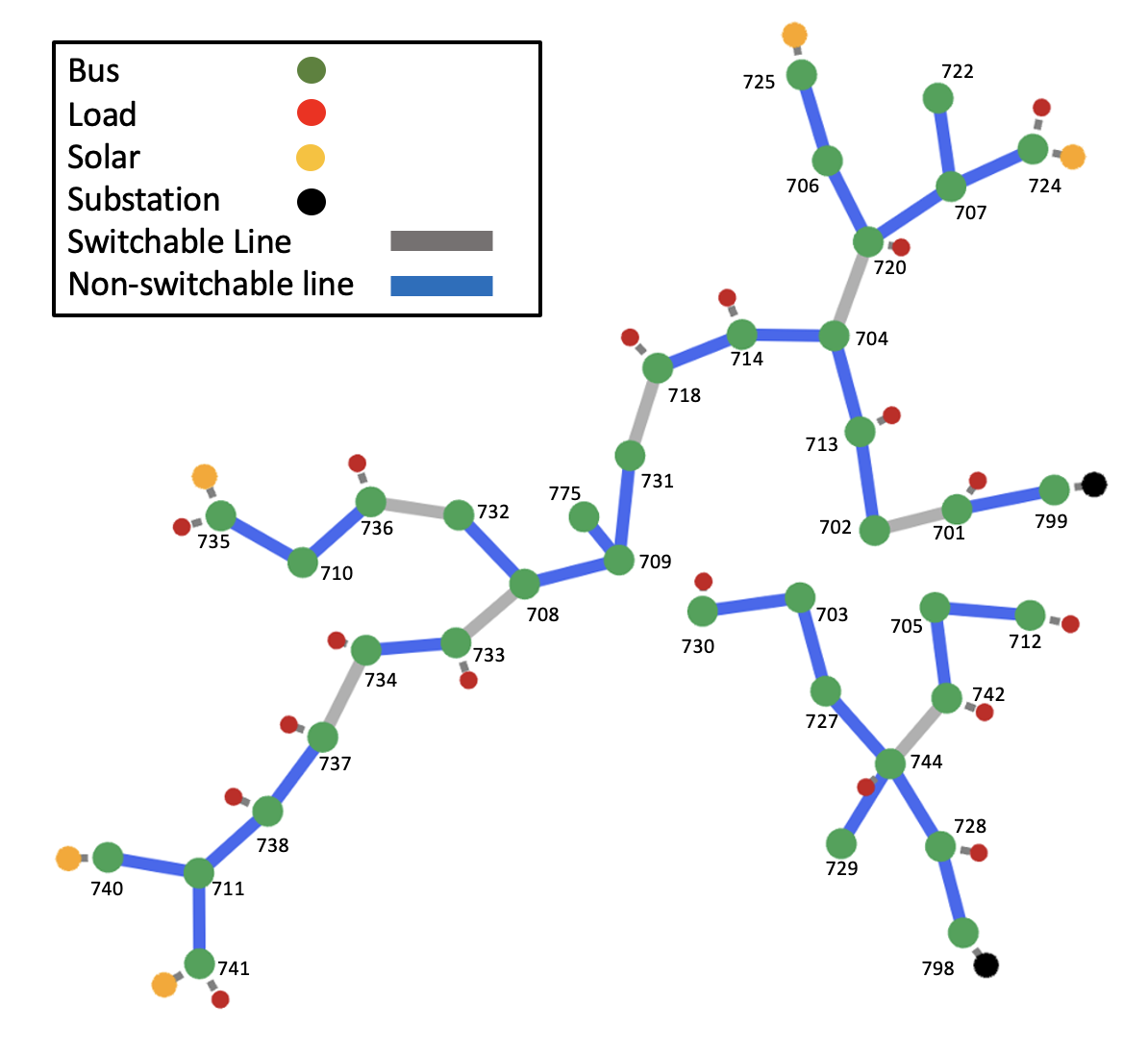} } 
\hspace{0.15cm}
   \subfloat[Scenario 2]{
      \includegraphics[trim=0cm 0cm 0cm 0cm,clip=true, width=0.26\textwidth]{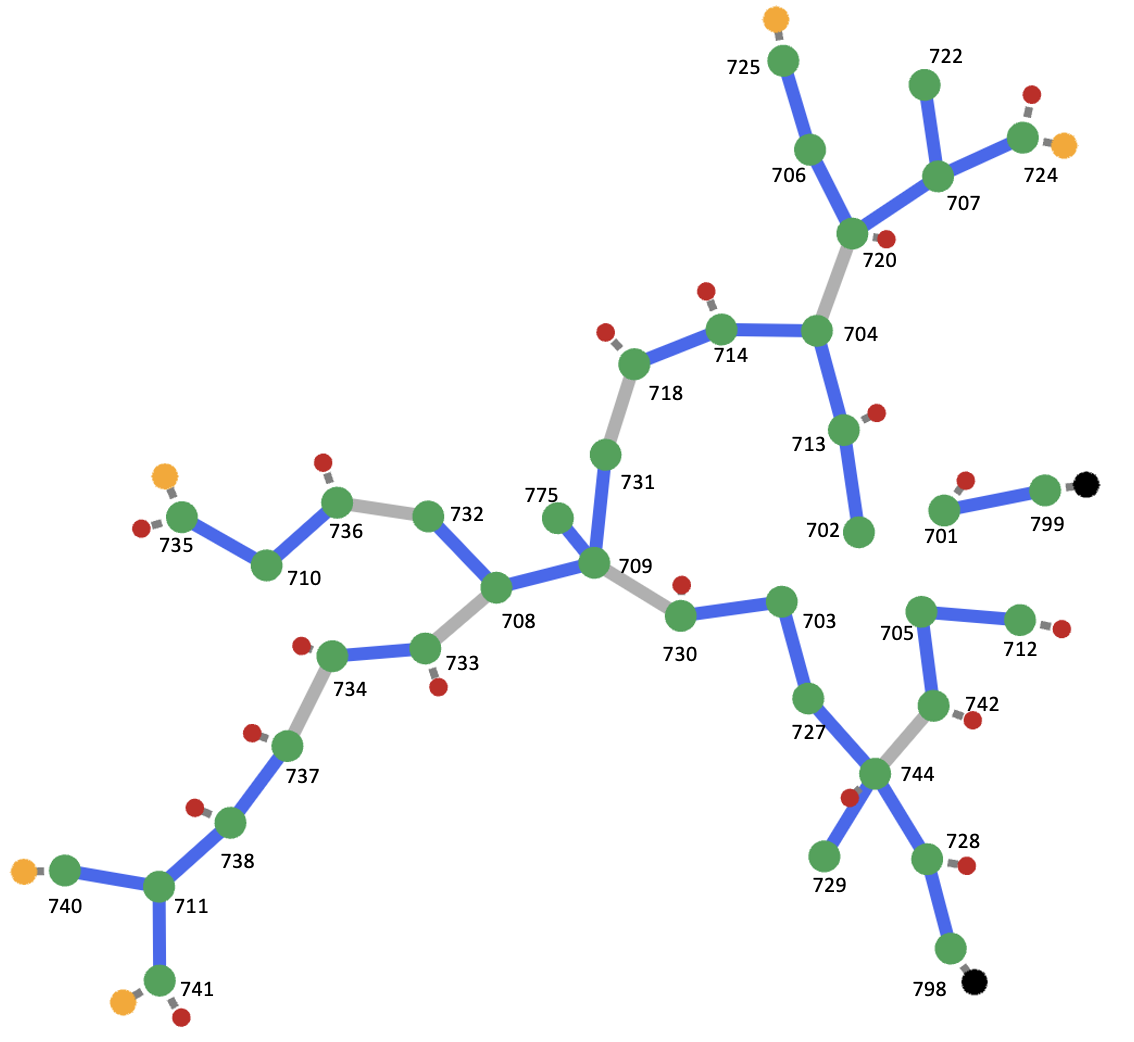} }
\hspace{0.15cm}
   \subfloat[Scenario 3]{
      \includegraphics[trim=0cm 0cm 0cm 0cm,clip=true, width=0.26\textwidth]{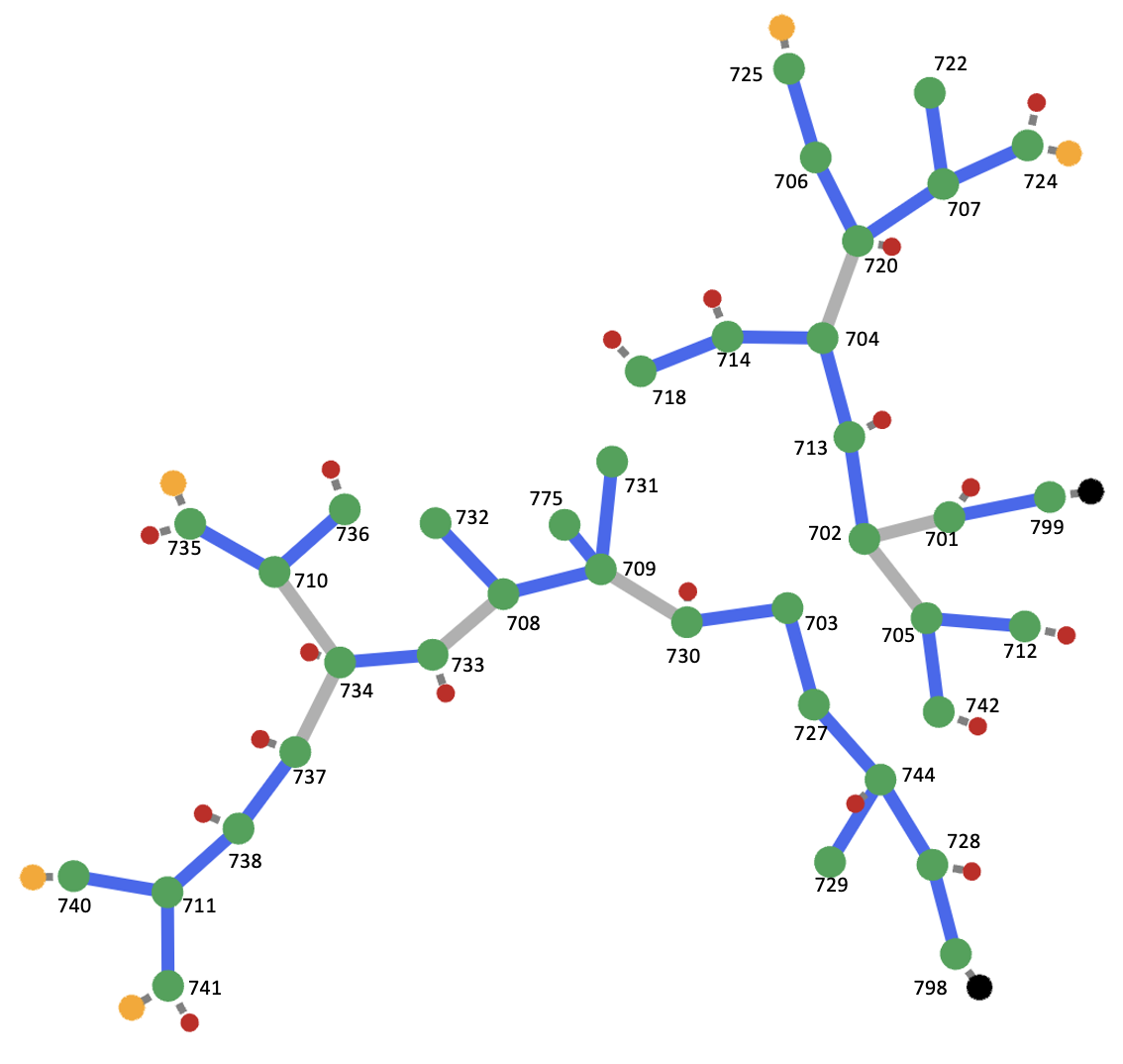}  }
\caption{Different network configuration outputs of the Neural Networks.}
\label{fig:different_configs}
\vspace*{-2mm}
\end{figure*}

\begin{table}[t!]
    \centering
    \caption{Summary of Deep Neural Network Settings}
    \begin{tabular}{|c|c|} \hline 
           \textbf{Item}&\textbf{Value}\\ \hline 
         No. of Hidden Layers&3\\ \hline 
         Activation functions used&ReLU, Sigmoid\\ \hline
         Learning Rate&0.0001\\ \hline 
         Optimizer&Adam\\ \hline 
         Batch Size&125\\ \hline 
         No. of epochs&250\\\hline
    \end{tabular}
    \label{tab:nn_settings}
\vspace*{-2mm}
\end{table}

\subsection{Neural Network Prediction Accuracy}
The neural network is designed to output the substation to which each load block is connected. Depending on a particular network, the outputs of some particular load blocks can be tied together. For instance, if load block $\mathbf{X}$ is connected to any substation through $\mathbf{Y}$, then the assignment of $\mathbf{X} \equiv \mathbf{Y}$. This can be applied to reduce the network complexity, enhance feasibility, and increase the model's overall accuracy. In this particular test network, we tied the substation connection of the load block connecting buses 731, 709, 775, 708, and 732 to all other load blocks lower down the network tree.

Applying the techniques explained in the paragraphs above, high accuracies were obtained after training the neural networks. The neural network was able to predict the substation to which each load block is connected with an accuracy of \textbf{\(\mathbf{95\%+}\)}.
Figure~\ref{fig:different_configs} shows the configurations that could be predicted based on the input load and PV profiles. The OPF for these networks can now be easily solved with the status of the switches fixed.

\subsection{Solving the OPF and time comparisons}
Upon successfully training the neural network, it can be used to predict the topology configuration for a given load profile and PV generation, after which the OPF calculation can be solved. We used the topology predictions for the 250 samples of testing data (5\% of the total training data) and performed the OPF calculations. The Table \ref{tab:opf_results} details the results from this calculation. 
The simulation results suggest that marginal loosening of the voltage bounds is required occasionally to attain feasible results.
It is also noted that a significant improvement in solve time was observed when we compared the two methods. In solving the offline problem, the average computation time for each iteration was $34.2$ seconds. This is insignificantly larger than the the computation time for the neural-network-based solution, which takes on average $1.54$ seconds. The reduction in computation time, which is more than an order of magnitude, could prove to be pivotal for the timely operation of the grid.
\begin{table}[t!]
    \centering
    \caption{Summary of OPF Solution Results}
    \begin{tabular}{|c|c|c|} 
        \hline 
        Voltage Bounds & No. of Feasible Outputs & Mean Solve Time (sec) \\ 
        \hline 
        0.9 - 1.1 & 80 / 250 & 2.31 \\ \hline
        0.875 - 1.125 & 210 / 250 & 1.59 \\ \hline
        0.85 - 1.15 & 230 / 250 & 1.52 \\ \hline
    \end{tabular}
    \label{tab:opf_results}
\end{table}


\section{Conclusion and Future Work}
The challenge of long computation time when solving the optimal reconfiguration problem in distribution networks of all scales can be easily overcome by harnessing the predictive ability of neural networks. In this paper, we proposed a learning approach to predict which substation supports each part of the distribution network. After learning this assignment, the problem is reduced to OPF problems with significantly fewer binary variables. These problems can then be solved much faster than the original reconfiguration problem. The approach was validated on the IEEE 37-bus distribution feeder, which was modified to include an additional substation and several distributed generation units. 
Future work will explore using graph neural networks to reduce learning complexity. We would also explore using equity matrices in computations and predictions to help ensure fairness.

\bibliographystyle{IEEEtran}
\bibliography{refs}

\vspace{12pt}

\end{document}